\documentclass[
reprint,
aip, 
jcp,
citeautoscript,
amsmath,amssymb,
longbibliography
]{revtex4-1}

\usepackage{graphicx}
\usepackage{dcolumn}

\usepackage{color}
\newcommand{\mos}{{MoS$_2$}}

\RequirePackage[xcolornames,svgnames,dvipsnames,rgb]{xcolor}
\pdfpageattr {/Group << /S /Transparency /I true /CS /DeviceRGB>>}

\begin{document}

\title{Adsorption of common solvent molecules on graphene and MoS$_2$ from first-principles}

\author{Urvesh Patil}
\author{Nuala M.~Caffrey}
\affiliation{School of Physics and CRANN, Trinity College, Dublin 2, Ireland}

\date{\today}

\begin{abstract}

Solvents are an essential element in the production and processing of two-dimensional (2D) materials. For example, the liquid phase exfoliation of layered materials requires a solvent to prevent the resulting monolayers from re-aggregating, while solutions of functional atoms and molecules are routinely used to modify the properties of the layers. It is generally assumed that these solvents do not interact strongly with the layer and so their effects can be neglected. Yet experimental evidence has suggested that explicit atomic-scale interactions between the solvent and layered material may play a crucial role in exfoliation and cause unintended electronic changes in the layer. Little is known about the precise nature of the interaction between the solvent molecules and the 2D layer. Here, we use density functional theory calculations to determine the adsorption configuration and binding energy of a variety of common solvent molecules, both polar and non-polar, on two of the most popular 2D materials, namely graphene and \mos. We show that these molecules are physisorbed on the surface with negligible charge transferred between them. We find that the adsorption strength of the different molecules is independent of the polar nature of the solvent. However, we show the molecules induce a significant charge rearrangement at the interface after adsorption as a result of polar bonds in the molecule.

\end{abstract}

\maketitle

\section{Introduction}

Two dimensional (2D) layered materials have attracted considerable attention since the discovery of graphene, due to their potential for advanced technological applications \cite{2Dgas,chhowalla2016two,AENM:AENM201301380,doi:10.1021/nl202675f}. 
Yet before they can be incorporated into devices, fabrication on a cost-effective, industrial scale must be achievable. 
There are two general approaches to the production of isolated 2D nanolayers, namely `top-down' and `bottom-up'. Bottom-up methods comprise of those which synthesize the layered material from atomic or molecular precursors and includes chemical vapor deposition (CVD). While offering a high degree of atomic control, bottom-up methods have generally a prohibitively high cost~\cite{cvd1,cvd2}. 
Top-down methods involve the extraction of individual layers from a parent layered crystal. An example of this is the isolation of a graphene monolayer from graphite by micro-mechanical cleavage. While cleavage techniques have been optimized to yield high quality 2D layers, they have a relatively low yield~\cite{micromechanical1}. In contrast, the liquid phase exfoliation (LPE) of layered materials is a scalable top-down method, capable of producing industrial quantities of monolayers at a low cost~\cite{lpe1,gLPE}.
Large shear forces, introduced in the presence of a solvent through either sonication, high-shear mixing or wet-ball milling, are used to overcome the van der Waal interactions binding the layers together~\cite{LPE}. The solvent then stabilizes the resulting nanosheets, preventing their aggregation or precipitation. Sheets with lateral sizes as large as 5~$\mu$m have been produced using this method \cite{paton2014scalable,KHAN2012470, o2012preparation, mLPE}.

The effectiveness of LPE is critically dependent on the choice of solvent \cite{gLPE,mLPE}. The simplified rule-of-thumb for solvation -- that polar solvents dissolve polar solutes and non-polar solvents dissolve non-polar solutes -- is no longer applicable. It was shown that matching the cohesive energies of the solute and solvent via the Hildebrand or Hansen solubility parameters can be a useful guiding principle in the search for an optimal solvent\cite{bergin2009multicomponent}. Yet this principle cannot be applied universally; in some cases the yield can be very low despite an excellent match between solute and solvent. For example, cyclopentanone and dimethyl pthlate have very similar Hansen and Hildebrand parameters yet the former is one of the best solvents for the exfoliation of graphite while the the latter is one of the worst~\cite{hernandez2009measurement}.
The failure of these empirical solubility models suggests that considering only macroscopic solution thermodynamics is not sufficient to find good solvents \cite{shen2015liquid}. 

Instead, explicit structural and electronic interactions between the solvent molecules and the solute may play an important role.
It has been suggested that solvent molecules can act as a `wedge', prising the layers apart at the edges, thereby improving the efficiency of subsequent exfoliation attempts~\cite{xu2014liquid, sresht2015liquid}. 
Mutual interactions may also result in the confinement of the solvent molecules at the surface or in interlayer spaces, resulting in changes in the entropic contribution to exfoliation~\cite{arunachalam2018graphene,MACIEL2011244solvation,govind2016dominance,howard2004formation}.

As well as in LPE, solvents are used in a variety of different material processing and purification tasks \cite{cullen2017ionic,torrisi2012inkjet,withers2014heterostructures}. 
In many cases, completely removing the solvent afterwards can be difficult. For example, N-Methyl-2-pyrrolidone (NMP) is a typical solvent used in LPE and in other solvent processing tasks, but due to its high boiling point (202$^\circ$C) it can remain a persistent residue \cite{nguyen2014investigation}. 
It is generally assumed that such solvent molecules interact only weakly with the layered materials and so their effects can be neglected. However, this is not always the case and given the atomic thickness and large surface area of 2D layers, there may be unintended effects on the structural and electronic properties of the layer.  For example, Choi et al.~found that common solvents can transfer sufficient charge to transition metal dichalcogenide layers to cause measurable changes in their electrical and optical properties~\cite{Choi2016}.

Very little is known about the nature of the interaction between solvent molecules and 2D layered materials on the atomic level.
In this work, we use density functional theory to systematically determine the ground state adsorption configuration of a variety of solvent molecules on two of the most widely studied 2D materials, namely graphene and hexagonal \mos. We choose six representative solvents from the polar protic (2-propanol), polar aprotic (bendaldehyde, cyclopentanone and N-Methyl-2-pyrrolidone (NMP)) and non-polar (toluene and chloroform) solvent families. These are shown in Fig.~\ref{fig:mol}(a).
We determine their adsorption configuration and binding energy and show that these molecules are physisorbed on the surface with little charge transfer between the two. Despite this, a significant charge rearrangement occurs at the interface due to an induced dipole interaction.

\section{Computational Methods}

\subsection{Density Functional Theory}

Density functional theory (DFT) calculations are performed using the projected augmented wave (PAW) method as implemented the {\sc vasp} code \cite{vasp1,vasp2,vasp3,vasp4}. The Perdew-Burke-Ernzerhof (PBE-PAW) potentials \cite{vaspPAW1,vaspPAW2} provided with the package are used. 
The optimized optB86b-vdW functional \cite{vaspvdw1,vaspvdw2,vaspvdw3,vaspvdw4,vaspvdw5} is used to approximate the exchange-correlation functional and to account for van der Waals (vdW) interactions. This functional was previously shown to provide accurate binding energies for both gas phase clusters and bulk solids and for molecular adsorption on transition metal surfaces\cite{carrasco2014insight}.

\begin{figure}
 \includegraphics[width = \columnwidth]{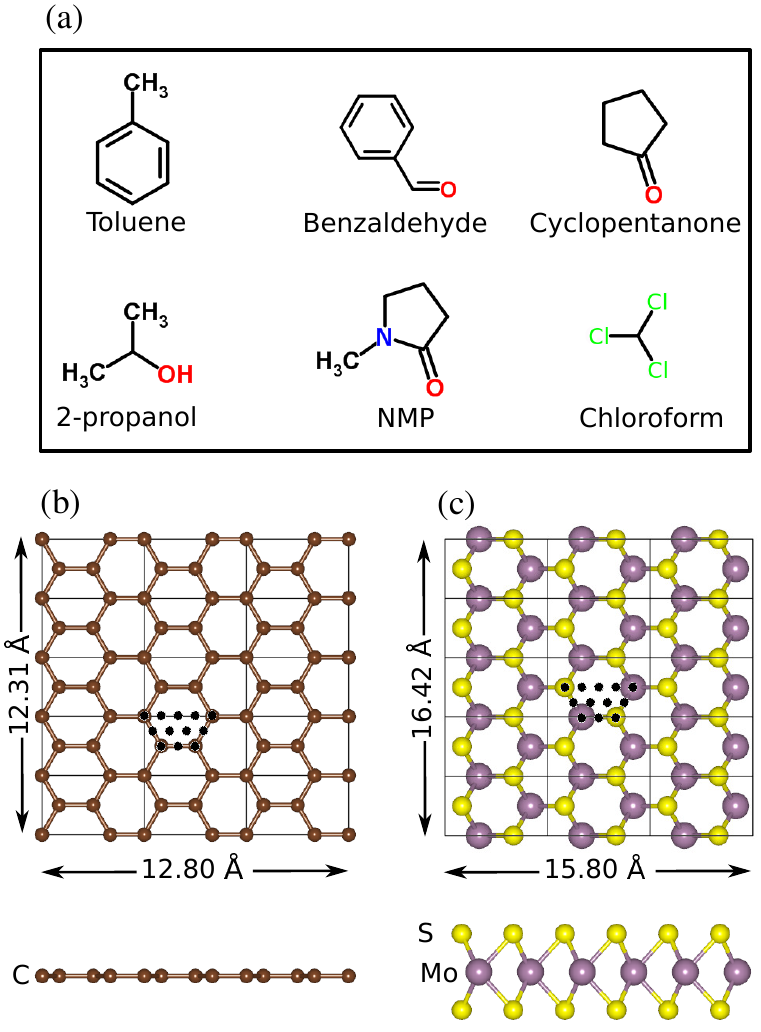}
 \caption{\label{fig:mol}  (a) Geometry of the six solvent molecules considered in this study. Top and side view of the (b) graphene and (c) \mos\ supercell used in this work. The 12 irreducible adsorption points are shown as black dots on the lattice.}
\end{figure}

In order to model the adsorption of isolated molecules, a $3 \times 5$ orthorhombic unit cell of both graphene and \mos\ is used, as shown in Fig.~\ref{fig:mol}(b) and (c). As a result, there is a minimum distance of at least 10~\AA\ between periodic images of the molecules. Furthermore, a  vacuum layer of at least 15~\AA\ is included in the direction normal to the surface to ensure no spurious interactions between repeating layers, and the dipole correction is applied.

The graphene (\mos) Brillouin zone is sampled with a $5 \times 5 \times 1 $ ($3 \times 3 \times 1 $) Monkhorst-Pack mesh~\cite{MP} to carry out structural relaxations to a force tolerance of $0.02$~eV/\AA. All atoms in the unit cell are allowed to move, including those of the substrate. The electronic properties are then calculated using a $k$-point sampling of $11 \times 11 \times 1$. In all cases, a plane wave cutoff of 500~eV is used to converge the basis set. 

The determination of charge transfer depends sensitively on how the charge density is assigned to each atom. Here, we use both the Density Derived Electrostatic and Chemical (DDEC) net atomic charges \cite{DDEC6} scheme as implemented in the chargemol program and the Bader partitioning scheme \cite{bader1,bader2}.

\subsection{Mapping the Configuration Space}

For multi-atom adsorbents, such as the molecules considered here, there is a large phase space of possible adsorption configurations. In order to find the lowest energy binding site, we follow a process similar to {\AA}kesson et al.~\cite{Akesson2012}, extended to include molecular rotations. Note that while the symmetry of the substrate is taken into account when creating the initial adsorption configurations, nothing is assumed about the molecular symmetry.

The following workflow is used to determine the ground state binding configuration:
The individual components, i.e., the molecule and the 2D layered material, are first relaxed to determine their isolated structures. 
A uniform grid is then defined at a typical binding height (3.5~\AA) above the surface of each material, as shown in Fig.~\ref{fig:mol}(b) and (c). The grid spacing is defined as $d/2$ where $d$ is the C--C or Mo--S bond length, projected in-plane. The center of mass of each molecule is placed at each grid point. 

Due to the low adsorption concentration considered here, each molecule will minimize its total energy by maximising its total area of overlap with the surface, i.e., planar molecules adsorb flat against the substrate \cite{C0CC02675A,Moses2009}. With this restriction, molecular rotations, in steps of $5^\circ$, around an axis normal to the basal plane of the substrate are considered. Out-of-plane rotations are also included.
Planar molecules such as benzaldehyde have only one indistinguishable out-of-plane rotation. NMP, cyclopentanone and toluene are non-planar with two possible rotational configurations obtained by a $180^\circ$ rotation out-of-plane. Chloroform has four possible rotational configurations: two in which the H--C bond is perpendicular to the plane of graphene, and another two in which the H--C bond is at $60^\circ$ to the plane. Finally, 2-propanol also has four possible rotational configurations: two orientations in which the C--O bond is perpendicular to the surface and another two in which it is parallel.
A structure matching algorithm, as implemented in pymatgen~\cite{pymatgen}, then reduces the total number of configurations.

The total energy of each of these configurations, without relaxation, is calculated. The entire procedure is then repeated for a sub-set of these configurations at a lower height in steps of 0.25~\AA\ until the lowest energy adsorption height is found. At this stage, a structural optimization of all structures at local minima with total energies within 0.05~eV of the global minimum is performed. The configuratation with the lowest total energy after this structural optimization is the ground state configuration.

\section{Results}

\subsection{Ground State Configurations}
\begin{figure}[ht!]
	\centering		
		\includegraphics[width=\columnwidth]{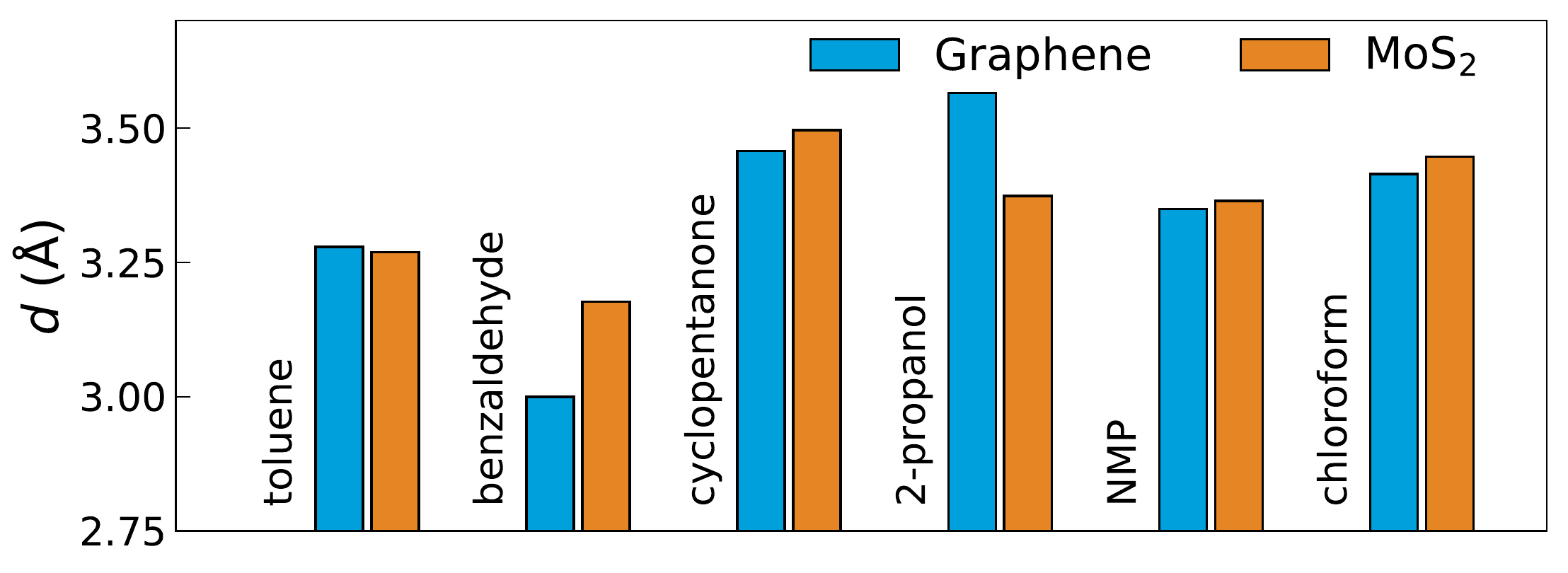}
	\caption{\label{fig:BH}The binding height of center of mass of the molecule from the basal plane of the substrate.}
\end{figure}

The solvent molecules are found to adsorb at an average binding height of 3.35~\AA\ from the surface of both graphene and \mos. The binding heights are shown in Fig.~\ref{fig:BH}. The smallest binding height is found for benzaldehyde on graphene (3.00~\AA), while the largest binding height of 3.56~\AA\ is found for 2-propanol on graphene.  These heights are consistent with physisorption \cite{BORCK2017162,Moses2009,Akesson2012}.

\begin{figure*}[ht!]
	\includegraphics{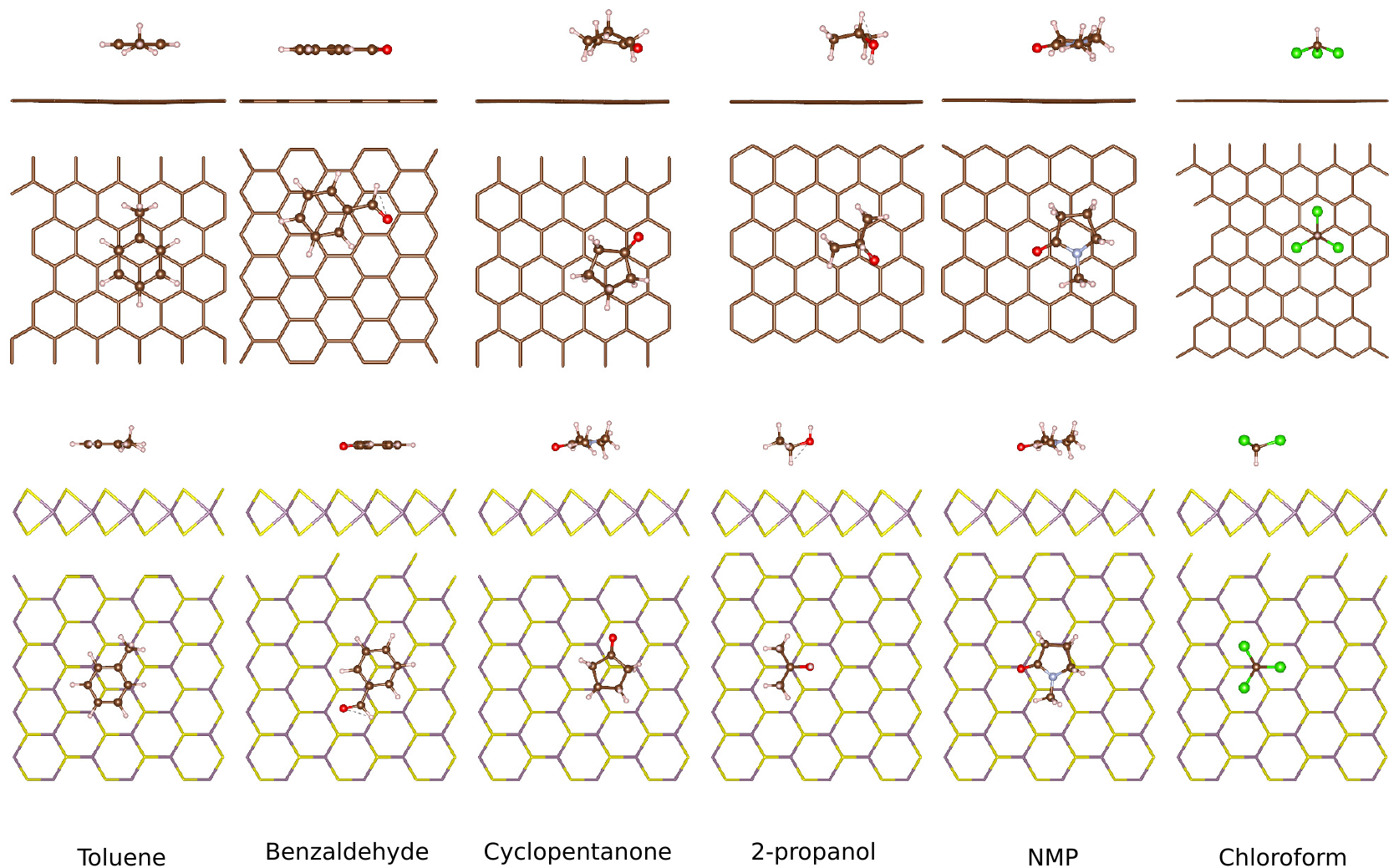}
	\caption{\label{fig:gpos}Top (bottom) panel: Side and top view of the lowest energy binding site of the solvent molecule on graphene (\mos).}
\end{figure*}

The geometrically optimized minimum energy configurations for each of the six solvent molecules adsorbed on graphene are shown in the top panel of Fig.~\ref{fig:gpos}. Molecules which contain a six-member ring are found to adsorb such that every alternate atom of the carbon ring is on top of a carbon atom in the graphene sheet, similar to the AB-stacking of two adjacent carbon layers in a graphite crystal~\cite{graphiteAB}. This is particularly evident for toluene and benzaldehyde where small deviations from the perfect AB-type stacking are dictated by the functional group attached to the ring. 
The methyl group of toluene is adsorbed at a `top' position, i.e., on top of a graphene carbon atom, with the edge of the methyl group tripod facing the graphene lattice, in agreement with Borck et al.\cite{borck2017methylbenzenes} In contrast, the functional aldehyde (CHO) group of benzaldehyde is adsorbed at a hollow position.
This is due to the different hybridizations of the carbon atoms in the two functional groups -- the carbon atom in the methyl group is $sp_3$ hybridized, whereas it is $sp_2$ hybridized in the CHO group. 
As the aldehyde oxygen atom has a partial negative charge, it prefers to adsorb close to a graphene bridge site. 

The carbon atoms in cyclopentanone are $sp_3$ hybridized with the exception of that bonded to oxygen, which is $sp_2$ hybridized. The three carbon atoms bonded to hydrogen atoms which point towards the graphene layer are located above hollow sites. The carbon atom bonded to a hydrogen atom which points away from the graphene layer is adsorbed above a carbon top site. The remaining electropositive carbon atom is adsorbed on a graphene top site, while the electronegative oxygen atom is located close to a graphene bridge site.
Similarly, the oxygen atom in 2-propanol adsorbs close to a bridge site and all $sp_3$ hybridized carbon atoms avoid the top sites. It maximizes its surface contact area by adsorbing such that the C--O bond is approximately parallel to the graphene layer.

For the case of NMP, the electronegative oxygen and nitrogen atoms dictate the orientation of adsorption by adsorbing close to bridge sites. Fixing the adsorption position of these two atoms determines the orientation of the rest of the molecule.
Finally, for the case of chloroform, each of the chlorine atoms adsorbs close to a hollow site, with the hydrogen pointing away from the layer in a so-called ``H-up'' configuration. Note that this is a different adsorption configuration to that found by {\AA}kesson et al.~due to the more restrictive configuration space considered in that work~\cite{Akesson2012}.
In all cases the deformation in graphene substrate is less than 0.1~\AA. 

The geometrically optimized configurations of the molecules adsorbed on \mos\ are shown in the bottom panel of Fig.~\ref{fig:gpos}. In all cases, molecules with hydrogen atoms which point towards the \mos\ surface prefer to adsorb such that they are located in the hollow formed by the sulfur atoms, i.e., directly on top of the metal atoms. 
For benzaldehyde and toluene, the carbon ring prefers to have alternate carbon atoms above the metal atoms with the center of the ring directly above a sulfur atom. Similarly, for cyclopentanone, the center of the carbon ring prefers to adsorb directly above a sulfur atom with the carbon atoms located either directly on top of the molybdenum atoms or in the hollow of the substrate hexagon. 2-propanol occupies the valley created by the sulfur atoms, with the functionalized carbon atom located on top of the metal atom. Note that this is a 180$^\circ$ out-of-plane rotation with respect to the orientation of the same molecule on graphene. For the case of NMP, the electronegative oxygen atom is adsorbed on top of the metal atom with the orientation of the rest of the molecule dictated by the hydrogen atoms which point towards the surface.

Finally, the hydrogen atom of chloroform also prefers to adsorb in the valley created by the sulfur atoms, directly above the metal atom, so that the molecule is in a ``H-down'' configuration. This is in contrast to its binding configuration on graphene where it adsorbs with the hydrogen atom pointing away from the surface, i.e.~``H-up''.
In all cases the deformation of \mos\ substrate after solvent adsorption is negligible.

\subsection{Binding Energy}

\begin{figure}[ht!]
	\centering
		\includegraphics[width=\columnwidth]{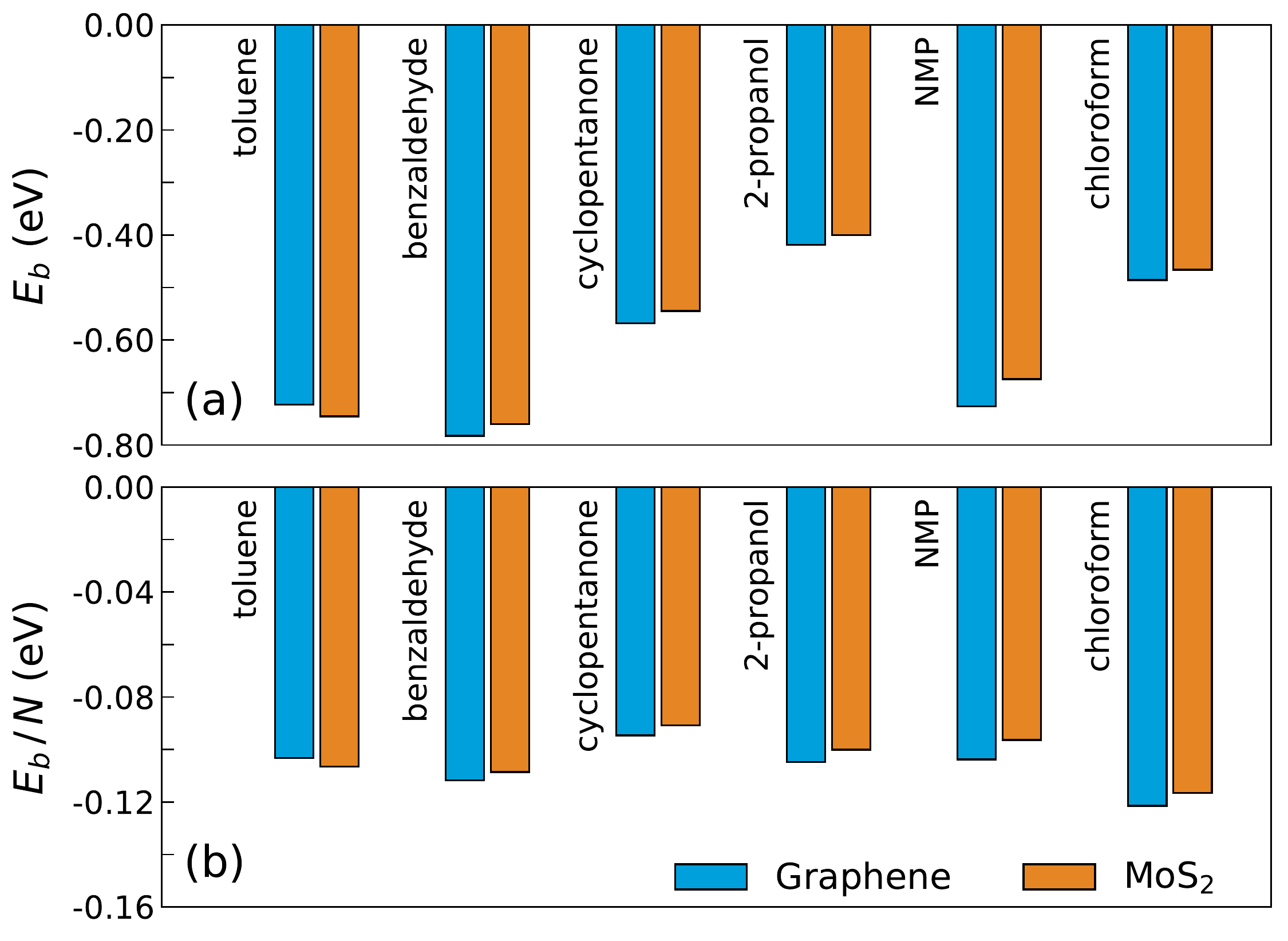}
	\caption{\label{fig:be}(a) Binding energy of each molecule on a graphene (blue) and \mos~(orange) substrate. (b) Binding energies normalized by the number of non hydrogen atoms, $N$, in the molecule.}
\end{figure}

The binding energy between the layered material and the adsorbed solvent molecule is defined as: 
$$E_b = E_{\mathrm{mol+layer}}-E_{\mathrm{layer}}-E_{\mathrm{mol}}$$
where $E_{\mathrm{layer}}$ is the total energy of the clean monolayer, $E_{\mathrm{mol}}$ is the total energy of the isolated molecule and $E_{\mathrm{mol+layer}}$ is the total energy of the combined system.
The binding energies of each solvent molecule adsorbed on both graphene and \mos\ are shown in Fig.~\ref{fig:be}(a). They range between -0.4~eV and -0.79~eV per molecule. The binding energy of each molecule differs by no more than 7\% when adsorbed on graphene compared to \mos. The molecular binding energy rescaled by the total number of atoms in that molecule, excluding hydrogen, $N$, is then shown in Fig.~\ref{fig:be}(b). In all cases, the normalized binding energies lie in a narrow range between approximately 90 and 120~meV/atom, and with a difference of no more than 5~meV/atom between individual molecules adsorbed on graphene and \mos. 
A similarly narrow range of normalized binding energy was found for aromatic and conjugated compounds adsorbed on \mos\ \cite{Moses2009} and graphene \cite{Chakarova-Kack2010} and shown experimentally for acenes adsorbed on copper surfaces~\cite{linearcopper}. This is evidence of the dominance of the van der Waals contribution to the binding energy and is supported by the fact that a positive binding energy is found for all molecules adsorbed on both substrates when the contribution from the vdW correction is excluded.

\subsection{Charge Transfer and Rearrangement}

The magnitude of total charge transfer between the molecules and both graphene and \mos\ is no more than $0.11e^-$ per molecule as determined by both the Bader and the DDEC methods. In some cases, these two methods do not agree on the direction of the charge transfer. Given the difficulties in partitioning space in order to assign charge to the molecule or substrate, this magnitude of the charge transfer may be considered essentially zero.

\begin{figure}[ht!]
	\centering		
		\includegraphics[width=\columnwidth]{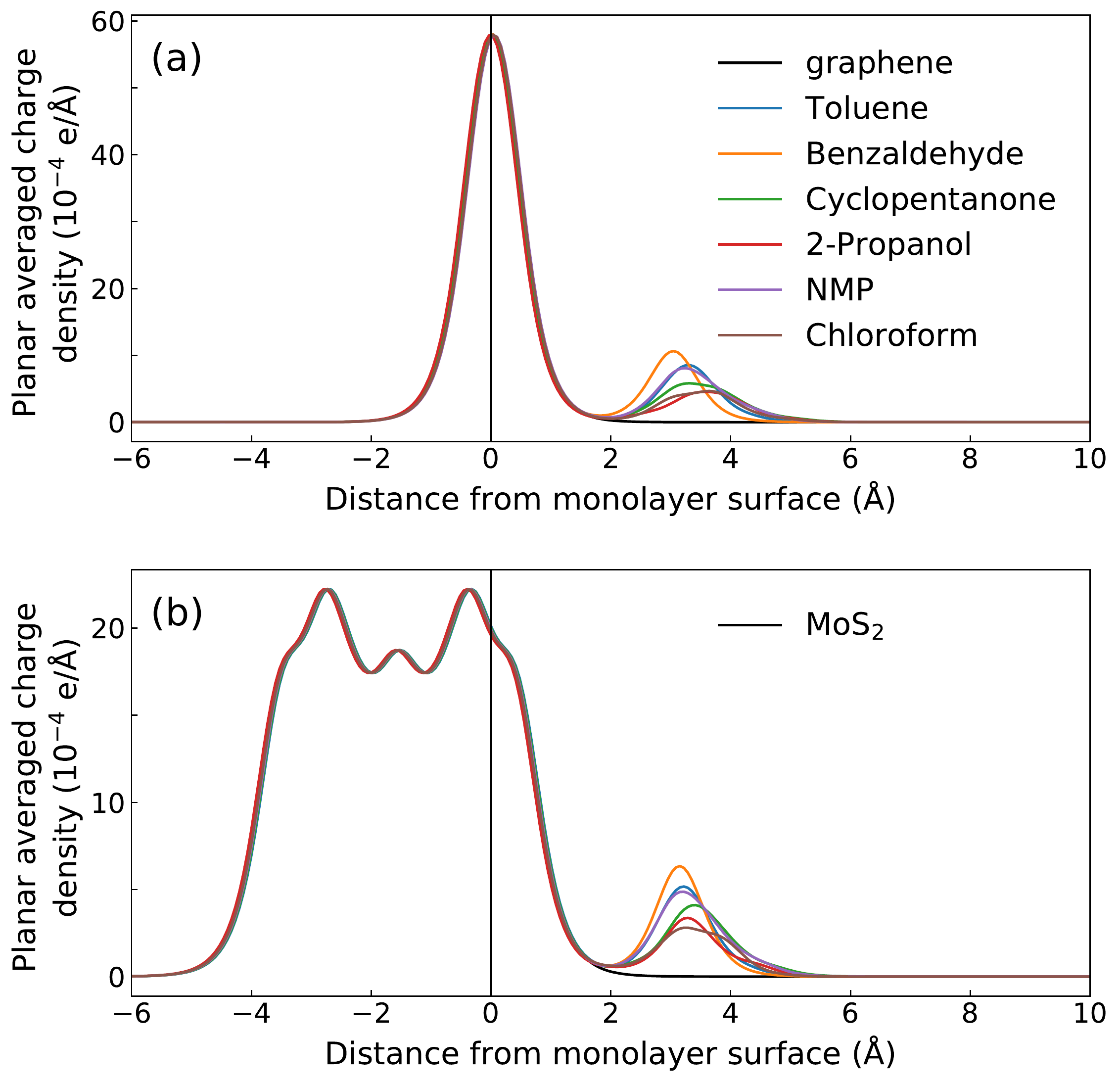}
	\caption{\label{fig:avg} Planar average of the charge density of each of the six molecules adsorbed on (a) graphene and (b) \mos\ compared to the pristine monolayers. The vertical lines indicate the positions of atoms in the monolayers.}
\end{figure}

Fig.~\ref{fig:avg} shows that there is a negligible difference in the charge density located on the monolayers before and after adsorption. From this we can conclude that the solvent molecules are physisorbed on the 2D layers. 

\begin{figure*}[!ht]
	\includegraphics{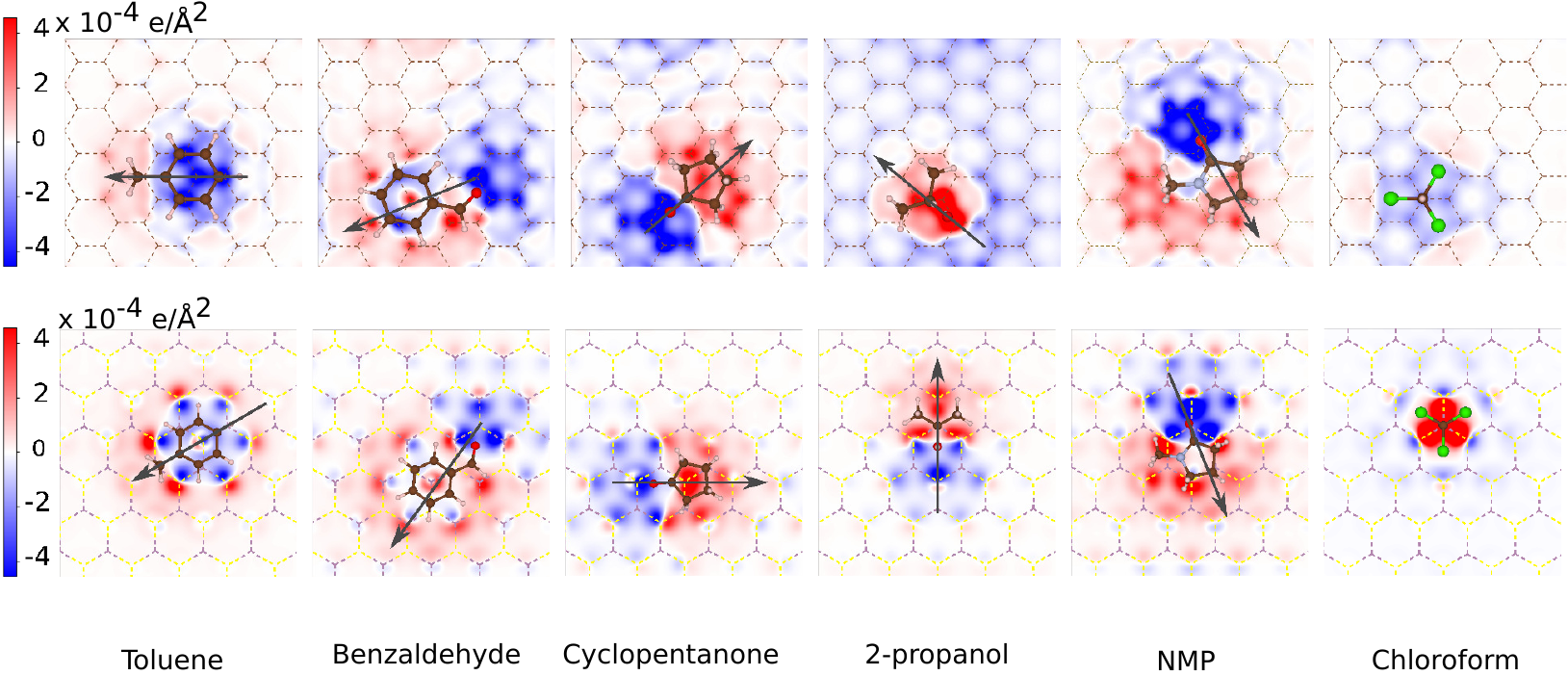}
	\caption{\,Top (bottom) panel: A slice through the charge density difference 0.5~\AA\ above the graphene (\mos) plane. Blue represents electron density depletion and red represents an electron density accumulation. The arrows represent the in-plane direction (but not magnitude) of the molecular dipole. The dipole of chloroform is perpendicular to the plane of the monolayer, pointing towards (away from) the layer for the case of graphene (\mos).}\label{fig:gmct}
\end{figure*}

Notwithstanding the negligible charge transfer involved, significant charge reorganization occurs on both the molecule and the 2D layer as as result of their interaction. To visualize this, we show in Fig.~\ref{fig:gmct} a slice through the charge density difference that occurs after molecular adsorption on graphene and \mos\ at a height of 0.5~\AA\ above the plane of the monolayer. The charge density difference is defined as: 
$$\Delta \rho = \rho_{\mathrm{mol+layer}} - \rho_{\mathrm{mol}} -\rho_{\mathrm{layer}}$$ 
where $\rho_{\mathrm{mol+layer}}$, $\rho_{\mathrm{mol}}$ and $\rho_{\mathrm{layer}}$ are the charge densities of the molecule adsorbed system, the isolated molecule and the isolated layer, respectively. A charge rearrangement reminiscent of image charges\cite{imgChg} on a metal is found to occur after molecular adsorption on graphene. As a result of their high polarizabilities\cite{santos2015electric}, the substrate's charge density is modified by the polar bonds of the adsorbing molecule. This can be seen as the response of the layer to the net dipole of the molecule. The molecule then interacts with its image charge. 

For the case of toluene, the small net molecular dipole points towards the methyl group. As a result, a small charge accumulation (red) is evident beneath the methyl group and a charge depletion (blue) occurs beneath the carbon ring. 
This dependence of the charge rearrangement on the molecular dipole is particularly evident for molecules with an electronegative oxygen atom, such as benzaldehyde, cyclopentanone and NMP.
In these cases, charge depletion occurs beneath the oxygen atom, whereas there is charge accumulation beneath the carbon ring. This is true for those molecules adsorbed on both graphene and \mos. 
Similarly, in 2-propanol the net dipole points away from the oxygen atom. However, the response of the 2D layer to 2-propanol depends on the out-of-plane rotation of the molecule. For the case of graphene, the molecule is adsorbed with the hydrogen atom, which is bound to the oxygen atom, pointing towards the surface. This hydrogen atom has a partial positive charge, and so results in charge accumulation in the layer directly beneath it. When adsorbed on \mos, that hydrogen atom points away from the surface. The charge depletion in the sulfur atoms of the substrate is then as a result of the partial negative charge on the oxygen atom. 
Finally, for the case of chloroform adsorption, the net dipole is perpendicular to the layers so that the changes in charge density around the molecule are symmetric. As the chlorine atoms have partial negative charges, charge depletion is evident directly beneath them, whereas there is a charge accumulation beneath the hydrogen atom which adsorbs on top of the metal atom.

\section{Conclusion}

We have determined the adsorption configuration of six common solvent molecules on the basal plane of both graphene and \mos\ using first-principles calculations which take van der Waal interactions into account. The calculated binding energies, adsorption heights and charge transfer all show that the solvent molecules are physisorbed on graphene and \mos, with only minor variations in binding height and binding energy between the different molecules and on the two different monolayers. For those molecules which contain a carbon ring, the lowest energy adsorption configuration on graphene is one in which a Bernal-like stacking arrangement of the carbon atoms is achieved. Non-planar molecules which have hydrogen atoms pointing towards the surface adsorb such that those atoms are located in the hollow site of the substrate lattice. We find that the orientation of both 2-propanol and chloroform are rotated by 180$^\circ$ when comparing adsorption on graphene and \mos. Finally, despite negligible charge transfer between the solvent and monolayers, there is a significant charge rearrangement within the substrate layers in response to the partial charges on the atoms in the molecules, similar to the creation of an image charge in metals.

\begin{acknowledgments}
This work was supported by a Science Foundation Ireland Starting Investigator Research Grant (15/SIRG/3314). Computational resources were provided by the supercomputer facilities at the Trinity Center for High Performance Computing (TCHPC) and at the Irish Center for High-End Computing (project tcphy091b, tcphy084c). The authors would like to thank Dr.~Thomas Archer for providing the ``gollum'' script, a modified version of which was used to perform the high-throughput calculations. 
\end{acknowledgments}

\end{document}